\documentstyle[epsfig,aps]{revtex}
\begin{document}
\draft
\onecolumn

\wideabs{
\title{Implications of a $300-500~GeV/c^2$ $Z'$ boson
on $p\bar{p}$ collider data at $\sqrt{s}=1.8$~TeV}
\author{Arie~Bodek$^{1}$ and Ulrich~Baur$^{2}$
    (UR-1617,UB-HET-01-02)
    }
\address{$^{1}$Department of Physics and Astronomy,
    University of Rochester, Rochester, NY 14627\\
    $^{2}$Department of Physics, State University of New York at Buffalo,
Buffalo, New York, 14260
    }

\maketitle
\begin{abstract}
Recent analyses of precision low energy
electroweak data indicate that the deviation from the Standard Model
predictions  of the measurement of  atomic
parity violation ($2.3\,\sigma$), the  effective number
of massless neutrinos ($2\,\sigma$), and $A_b$ ($2.7\,\sigma$) could
be better described if the existence of an extra $Z'$ neutral gauge boson
is assumed. We investigate the implications of a
300~--~500~$GeV/c^2$ extra $Z'$
on current $p\bar{p}$ collider data at $\sqrt{s}=1.8$~TeV, including
the forward-backward charge asymmetry
for very high mass $e^+e^-$ pairs, and the invariant
mass spectrum of
high mass $e^+e^-$, $\mu^+\mu^-$, $t\overline{t}$ and  $b\overline{b}$
pairs.  For example,
a 500~$GeV/c^2$ $Z'$ with a total production cross section
of $\approx 3\,pb$
and enhanced coupling to the third generation,
    better describes both the low energy and the Tevatron data.
\end{abstract}
\pacs{PACS numbers:  12.14.Ji, 12.15.Mm, 12.38.Qk, 12.60.-i,12.60.Cn}
\twocolumn
}
%
%
%
Recent analyses~\cite{zfitrosner,zfit} of precision
electroweak data indicate that there are several measurements
for which the
deviations from the Standard Model predictions
are larger than two standard deviations  ($\sigma$).
These include the measurements of  atomic
parity violation~\cite{atomic} ($2.3\,\sigma$), the effective number
of massless neutrinos~\cite{LEP} ($2\,\sigma$),
and $A_b$~\cite{LEP} ($2.7\,\sigma$). These analyses show that
the data are better described if an extra $Z'$ neutral gauge boson
is assumed.  In this analysis, we investigate if there is
evidence for
a $Z'$ boson in current $p\bar{p}$ collider data at
$\sqrt{s}=1.8$~TeV.

We take the parameters and couplings of the
$Z'$ to the first generation quarks and leptons from the analysis of the atomic
parity violation data by Rosner~\cite{zfitrosner}. Rosner's
analysis indicates that the atomic parity violation
data~\cite{atomic} are better described with an
$E_6$ extra $Z'$ boson. Within this model~\cite{E6} there is a continuum of
$Z'$ possibilities given by  $Z' =
Z_{\psi}\cos{\phi}+Z_{\chi}\sin{\phi}$. Rosner's
analysis of the atomic parity violation
data yields a best fit for a region
of allowed $Z'$ mass, $M_{Z'}$, and $\phi$. For example,
For $\phi=120^0$ the data are best fit with an
    $E_6$ $Z'$ with a mass of about 800~$GeV/c^2$. For
an $E_6$ $Z'$ with $M_{Z'}=500~GeV/c^2$,
the atomic parity violation data are best fit with $\phi=70^0\pm5^{0}$ and
$\phi=160^0\pm5^{0}$, and for a $Z'$ with a mass of 350~$GeV/c^2$,
these data are best fit with $\phi=60^0\pm5^{0}$ and
$\phi=173^0\pm5^{0}$.

Erler and Langacker~\cite{zfit} extend the $Z'$ analysis to include all
precision electroweak data, and include more general classes
of $Z'$ models. In one of the cases, the
analysis is extended to  allow for the coupling to the third generation to be
different from the coupling to the first two generations. With these
additional parameters, they are not able to place a constraint on
$M_{Z'}$, but the low energy electroweak data prefer a $Z'$ with
a small (but finite) mixing to the $Z$, and a
large coupling to the third generation,
    as expected in some models~\cite{zfit,topc}.
The larger coupling to  $b\overline{b}$  pairs
    is needed
to account for the  $2.7\,\sigma$ deviation of  $A_b$~\cite{LEP}  from the
Standard Model prediction.

Although the mass limits~\cite{zprime} from CDF
and \mbox{DO\hspace*{-1.5ex}/}~\cite{D0}
for a variety of $Z'$ models are in the 600~$GeV/c^2$ range, the limits are
reduced~\cite{zprime}
by 100 to 150~$GeV/c^2$, if the $Z'$ width (typically
$\Gamma_{Z'}\approx 0.01\, M_{Z'}$)
is increased to  account for the possibility of
additional decays modes to
exotic fermions (which are predicted in $E_6$ models), and/or
supersymmetric particles. The limits are even lower if one includes
the possibility of a more general
model with enhanced couplings to the third generation.
Therefore,  we investigate the present Run~I collider data for
high mass $e^+e^-$, $\mu^+\mu^-$, $t\overline{t}$, and $b\overline{b}$
final states, and look for  possible
signatures for a $Z'$ extra gauge boson of the kind that is favored
by the low energy data. We constrain the relationship between
the couplings to the first two
generations and the mass of the $Z'$ to be
the same as that for an $E_6$ $Z'$ boson from Rosner's fits to
the low energy measurements.

In hadron-hadron collisions at high energies, massive $e^+e^-$ and
$\mu^+\mu^-$ pairs are produced via the Drell-Yan  $\gamma^*/Z$ process.
The presence of both vector
and axial-vector couplings in this process gives rise to an
asymmetry~\cite{E6},
$A_{FB}$, in the final-state angle of the $\it {lepton}$
in the rest frame of the
$e^+e^-$ and $\mu^+\mu^-$ pair (with respect to the $\it{proton}$
direction). Within the Standard Model,
for $M\gg M_Z$, the
large predicted asymmetry ($\approx 0.61$) is a consequence
of the interference between the propagators of the  $\gamma^*$ and $Z$.
New interactions
such as an extra $Z'$ boson result
in deviations from the standard model predictions
in both $d\sigma/dM$ and $A_{FB}$.

\begin{figure}
\begin{center}
\mbox{\epsfxsize=3.37in \epsfysize=3.55in
\epsffile{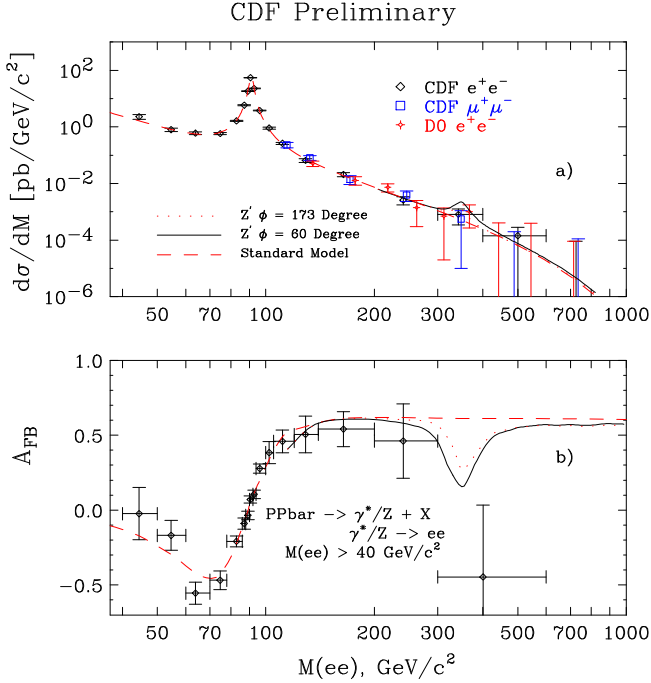}}
\end{center}
\caption{(a) $d\sigma/dM$
distribution of $e^+e^-$ (CDF and \mbox{DO\hspace*{-1.5ex}/})
and $\mu^+\mu^-$ pairs (CDF). The Standard Model
theoretical predictions (dashed line) have been normalized (by a factor
of 1.11) to the CDF data in the $Z$ boson mass region.
(b) CDF $A_{FB}$ versus mass compared to the
standard model expectation (dashed). Also shown are
the predicted theoretical curves ($\times 1.11$) for $d\sigma/dM$ and $A_{FB}$
with an extra $E_6$ boson  with $M_{Z'}=350~GeV/c^2$ and
$\Gamma_{Z'}=0.1\, M_{Z'}$, for $\phi=60^0$ (solid) and
$\phi=173^0$ (dotted).}
\label{fig1}
\end{figure}

\begin{figure}
\begin{center}
\mbox{\epsfxsize=3.37in \epsfysize=3.55in \epsffile{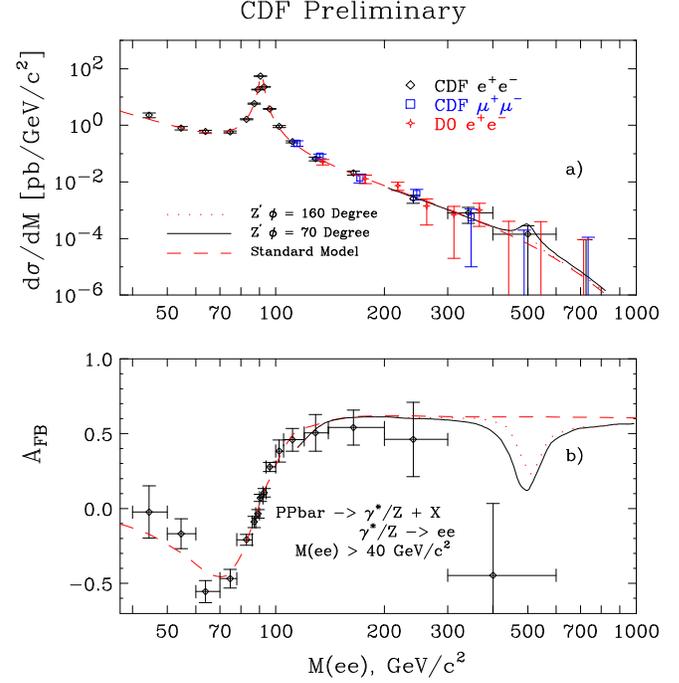}}
\end{center}
\caption{Same as Figure~\ref{fig1}, but shown here are
the predicted theoretical curves for $d\sigma/dM$ ($\times 1.11$) and $A_{FB}$
with an extra $E_6$ boson  with $M_{Z'}=500~GeV/c^2$ and
$\Gamma_{Z'}=0.1\, M_{Z'}$, for $\phi=70^0$ (solid) and $\phi=160^0$
(dotted).}
\label{fig2}
\end{figure}

Figures~\ref{fig1} and~\ref{fig2} compare the measured high mass Drell-Yan
$d\sigma/dM$ (CDF~\cite{DiMuPRD,asym} and
\mbox{DO\hspace*{-1.5ex}/}~\cite{D0}) and $A_{FB}$ (CDF~\cite{asym})
to theoretical predictions.
The Standard Model $d\sigma/dM$ curve is a QCD NNLO~\cite{NNLO} calculation
with MRST99 NLO PDFs~\cite{MRS}.
The predictions in Figure~\ref{fig1}(a) are normalized by a
factor $F=1.11$, the ratio of the CDF measured total cross section
in the $Z$ region~\cite{liu} to  the NNLO prediction (the
overall normalization uncertainties are 3.9$\%$ for the experimental
data and $5\%$ for the NNLO theory).
The Standard Model prediction for $A_{FB}$ has been calculated~\cite{baur}
in QCD-NLO with sin$^2\theta^{lept}_{eff}=0.232$. The measured
$d\sigma/dM$ and $A_{FB}$ values are in good agreement with the standard model
predictions.  In the two highest
mass bins (4 events in the 300~--~600~$GeV/c^2$ range),
$A_{FB}$ is about 2.2 standard
deviations  below the standard model prediction (there are 3 events in
the negative hemipshere and one event in the positive hemisphere).
An asymmetry in the 300~--~600~$GeV/c^2$ range which is smaller than the
Standard Model prediction
could result from the exchange of a 300~--~500~$~GeV/c^2$ $Z'$ gauge
boson.

For the $E_6$ $Z'$ models that were fit to the low
energy electroweak data, the couplings to the first
two generations are well constrained. Contributions to the $Z'$ cross
section from $t\bar t$ and $b\bar b$ annihilation are strongly
suppressed due to the small $t\bar t$ and $b\bar b$ parton luminosities, even
in the case where the $Z't\bar t$ and $Z'b\bar b$ couplings are strongly
enhanced. For a given $Z'$ mass and
$\phi$, the total $Z'$ production cross section (in all channels) in
$p\bar{p}$ collider data at $\sqrt{s}=1.8$~TeV thus
is determined by the couplings of the $E_6$ $Z'$ to the first and second
generation up and down type quarks.
The partial width to electrons is also determined.
The integrated cross section for $e^+e^-$ final
states, $\sigma\times$BR($Z'\to e^+e^-$), is determined by the
$Z'\to e^+e^-$ branching ratio, and
is  therefore proportional to $1/\Gamma_{Z'}$. In contrast, the
prediction for $A_{FB}$, which results from the interference between the
Standard Model and the $Z'$ amplitudes, is quite insensitive to the $Z'$
width.  In this study, we compare the Run~I collider data to a model
with a $Z'$ width of $\Gamma_{Z'}=0.1\, M_{Z'}$  (which is about a
factor of~10 larger than
the expected width for an $E_6$ $Z'$ boson for the case of universal couplings
to all three generations). This allows for enhanced couplings to
the quarks of the third generation, or for additional decay modes to exotic
fermions or supersymmetric particles.
The predictions for $d\sigma/dM$ and $A_{FB}$
for the case of the Drell-Yan process including an extra $E_6$ $Z'$
boson (for a given $M$ and $\phi$) are first calculated in leading order in
QCD (with MRSR2 NLO PDFs),
ignoring contributions from $t\bar t$ and $b\bar b$ annihilation.
$d\sigma/dM$ is then multiplied by a differential $K$-factor,
$K(M)$, which is determined by comparing the
leading order prediction for $d\sigma/dM$ including a $50~TeV/c^2$ $Z'$
(i.e. effectively no $Z'$ in the model) to the Standard
Model $d\sigma/dM$ prediction from the QCD NNLO~\cite{NNLO} calculation
(with MRST99 NLO PDFs).
In the 300~--~400 and 400~--~600~$GeV/c^2$
ranges, we obtain a differential $K$-factor of 1.172 and 1.154, respectively.
Since we plan to compare to CDF Drell-Yan data,
the predictions in Figure~\ref{fig1}(a) and~\ref{fig2}(a)
are normalized by a
factor $F=1.11$, the ratio of the CDF measured total cross section
in the $Z$ region~\cite{liu} to  the NNLO prediction.  Because the data
show a clustering of events at mass values of 350 and 500~$GeV/c^2$,
we investigate the signatures for a $Z'$ boson with these two mass
values. As mentioned earlier,
in all of the calculations we use a $Z'$ width of $\Gamma_{Z'}=0.1\,
M_{Z'}$.  If we
use a smaller width, the cross section in the di-lepton channel would
be larger by a factor proportional to $1/\Gamma_{Z'}$.

For an  $E_6$ gauge boson with $M_{Z'}=350~GeV/c^2$,
$\Gamma_{Z'}=0.1\, M_{Z'}$, and $\phi=60^0$ ($173^0$), the theoretical
prediction ($\times 1.11$) for the integrated total cross section
of  $e^+e^-$ pairs in  the $300-400~GeV/c^2$ mass range is 73 (31)~$fb$. The
corresponding cross section for the Standard Model prediction ($\times
1.11$) in this range is 54~$fb$.  Both the Standard Model cross section,
and the cross section including an additional 350~$GeV/c^2$ $Z'$ are
consistent with the observed CDF
cross section in this range of  $81\pm47~fb$.
Our predicted theoretical curves for $d\sigma/dM$ and $A_{FB}$
with an extra $E_6$ boson, for $M_{Z'}=350~GeV/c^2$ and
$\Gamma_{Z'}=0.1\, M_{Z'}$, are shown in Figure~\ref{fig1}.
The total production cross section ($\times 1.11$) for this
$Z'$ for $\phi=60^0$ ($173^0$) is 17 (11)~$pb$. For $A_{FB}$,
$\phi=60^0$ results in a better agreement of theory and data than
$\phi=173^0$. In the $300-600~GeV/c^2$ range, the probabilty that the
forward backward asymmetry predicted by the Standard Model agrees with
CDF data is 2.4\% (see Table~\ref{Table2}). For $M_{Z'}=350~GeV/c^2$,
$\Gamma_{Z'}=0.1\, M_{Z'}$, and $\phi=60^0$ ($173^0$), the corresponding
probability is 13.4\% (7.7\%).

Because the couplings of the $E_6$ $Z'$ to the first two generations of
quarks and leptons are constrained by the fit to the low energy
electroweak data, most of this cross section should appear in the form of
decay modes to either
exotic fermions (which are predicted in $E_6$ models) and/or
supersymmetric particles, or decays to third generation quarks.
Since the analysis of Erler and Langacker~\cite{zfit} indicates that a
larger coupling to the third generation is needed
to account for the $2.7\,\sigma$ deviation from the Standard Model of
$A_b$~\cite{LEP}, we look for $Z'$ signatures in the
invariant mass spectra of $t\overline{t}$ and $b\overline{b}$
high mass pairs.
    The CDF 95\% CL limit on the $b\bar b$ cross section~\cite{bbar}
    for a 350~$GeV/c^2$ $Z'$ varies from 12~$pb$ for a very
    narrow $Z'$ to 28~$pb$ for a $Z'$ with $\Gamma_{Z'}=0.3\, M_{Z'}$.
    Therefore, a production cross section of 11 to 17~$pb$
    for a 350~$GeV/c^2$ $Z'$ with $\Gamma_{Z'}=0.1\, M_{Z'}$ (which
predominantly
decays to  $b\overline{b}$ pairs) is consistent with these  $b\overline{b}$
limits. Note that $Z'\to t\bar t$ decays are either forbidden or
strongly phase space suppressed for $M_{Z'}=350~GeV/c^2$.

The Standard Model $e^+e^-$ Drell-Yan cross section for the $300-400~GeV/c^2$
mass range and the $Z'\to e^+e^-$ cross
section for $M_{Z'}=350~GeV/c^2$ and
$\Gamma_{Z'}=0.1\,M_{Z'}$, with $\phi=60^0$ and $\phi=173^0$ are
summarized in Table~\ref{Table1}. The
$Z'\to e^+e^-$ branching ratio and the total $Z'$ production cross
section are also shown in the table.

For an  $E_6$ gauge boson with $M_{Z'} = 500~GeV/c^2$,
$\Gamma_{Z'}=0.1\, M_{Z'}$, and $\phi=70^0$ ($160^0$), the theoretical
prediction ($\times 1.11$) for the integrated total cross section
of  $e^+e^-$ pairs in the $400-600~GeV/c^2$ mass range is 17 (7)~$fb$. The
corresponding cross section for the Standard Model prediction ($\times
1.11$) in this range is 17~$fb$.  Both the Standard Model cross section,
and the cross section including an additional $500~GeV/c^2$ $Z'$ are
consistent with the observed CDF
cross section in this range of  $28\pm28~fb$.
Our predicted theoretical curves for $d\sigma/dM$ and $A_{FB}$
with an extra $E_6$ boson, for $M_{Z'}=500~GeV/c^2$ and
$\Gamma_{Z'}=0.1\, M_{Z'}$, are shown in Figure~\ref{fig2}. The total
(with decay to all channels) production cross
section ($\times 1.11$) for this  $Z'$, together with the $Z'\to e^+e^-$
branching ratio and the Standard Model Drell-Yan cross section in the
$400-600~GeV/c^2$ mass range are listed in Table~\ref{Table1}. For
$A_{FB}$, the prediction for $\phi=70^0$ gives better agreement with the
data than that for $\phi=160^0$. For $\phi=70^0$ ($160^0$), the
probability that theory and data agree is 6.3\% (3.7\%) (see
Table~\ref{Table2}).

We now look for possible signatures of a $Z'$ with
these production cross sections in the  $b\overline{b}$ and
$t\overline{t}$  channels.
In the $b\overline{b}$ channel,
    the CDF 95\% CL cross section limit for a $Z'$ boson with 500~$GeV/c^2$
varies from 3.1~$pb$ for a very
    narrow $Z'$ to 5.5~$pb$ for a $Z'$ with $\Gamma_{Z'}=0.3\, M_{Z'}$.
    In the $t\overline{t}$ channel,
    the CDF 95\% CL cross section limit for a $Z'$ boson with
$M_{Z'}=500~GeV/c^2$ is 7.5~$pb$.
    Therefore, a production cross section of 1.7 to 3.2~$pb$
    for a 500~$GeV/c^2$ $Z'$ with $\Gamma_{Z'}=0.1\, M_{Z'}$ (see
Table~\ref{Table1}), which predominantly
    decays to  $b\overline{b}$ and/or $t\overline{t}$  pairs, is
    consistent with these limits.  It is interesting to note that
     the published $t\overline{t}$ and $b\overline{b}$ mass
    distributions show a slight excess of events in the 500~$GeV/c^2$
region.
\begin{figure}
\begin{center}
\mbox{\epsfxsize=3.00in \epsfysize=2.25in \epsffile{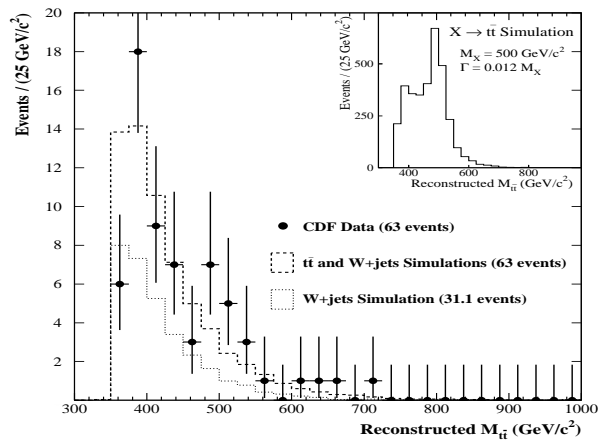}}
\end{center}
\caption{ The invariant mass
distribution of $t\overline{t}$ pairs at CDF (fit
with $m_t=175~GeV/c^2$).   The published CDF $Z'$
cross section 95\% CL upper limit in the  $t\overline{t}$
channel is 7.5~$pb$ at $M_{\bar tt}=500~GeV/c^2$. The
$1.4\,\sigma$ excess at 500~$GeV/c^2$ in the invariant mass spectrum
corresponds to a cross section of $2.3\pm1.7~pb$. }
\label{fig3}
\end{figure}

Figure~\ref{fig3} shows the CDF
published~\cite{tipton,cassada} invariant mass
distribution of $t\overline{t}$ pairs (assuming $m_t=175~GeV/c^2$). The
$1.4\,\sigma$ excess at 500~$GeV/c^2$ in the invariant mass spectrum
corresponds~\cite{cassada} to $\sigma\times$BR($t\overline{t})=2.3\pm1.7~pb$.
Therefore, the CDF $t\overline{t}$ data support the hypothesis of a $Z'$
boson with a large coupling to the third generation.
Although \mbox{DO\hspace*{-1.5ex}/}
has not searched for resonances in the $t\overline{t}$ channel,
the published \mbox{DO\hspace*{-1.5ex}/}
mass spectrum~\cite{Dzerott} for  $t\overline{t}$
events
is consistent with an enhancement in the $460-500~GeV/c^2$ region
at a similar level.
There is also small
$1\,\sigma$ excess at 500~$GeV/c^2$ in the CDF $b\bar b$ invariant mass
spectrum
which corresponds to  $\sigma\times$BR($b\overline{b})=1\pm 1~pb$.
Therefore, the CDF $b\overline{b}$ data are also consistent with the
hypothesis of a $Z'$ boson with a total production
cross section of 1.7 to 3.2~$pb$ which has a small branching
ratio to di-leptons and predominantly decays to top and bottom quarks.
Note, that for a $Z'$ that mixes with the $Z$, the fits to low energy
electroweak data at the $Z$ peak already constrain~\cite{zfit}
the level of the coupling of a $Z'$ to $\tau$ leptons  to be similar
to the couplings to electrons and muons. Therefore, an enhanced  signal in the
$\tau^+\tau^-$ channel is not expected for a $Z'$ which mixes with
the  $Z$.

In summary,  we find that
either a 350 or a 500~$GeV/c^2$ extra $Z'$ with $\Gamma_{Z'}=0.1\,
M_{Z'}$  and  enhanced couplings to the quarks in the third generation
not only gives a better description of the low energy electroweak data, but
also better describes the forward-backward asymmetry
for $e^+e^-$ pairs in the 300-600 $~GeV/c^2$ range.
A 500~$GeV/c^2$ extra $Z'$ also accounts for the
$1.4\,\sigma$ excess at 500~$GeV/c^2$ in the invariant mass spectrum of
high mass of $t\overline{t}$ events at CDF. A $Z'$ in the $350-500~GeV/c^2$
mass range is also compatible~\cite{cm} with the latest measurement of
the muon anomalous magnetic moment~\cite{mu}.
With the upgraded CDF and \mbox{DO\hspace*{-1.5ex}/}
detectors, and the anticipated factor of~20 higher luminosity in
Run~II, improved
searches for $Z'$ bosons could be made in the invariant mass spectrum
and in the forward-backward asymmetry for all three
di-lepton channels ($ee$, $\mu\mu$ and $\tau\tau$). In addition, the
improved silicon vertex detectors in both experiments will increase
the sensitivity of such searches in $t\overline{t}$ and $b\overline{b}$
final states.

One of us (UB) would like to thank the Fermilab Theory
Group, where part of this work was carried out, for its generous
hospitality. This work has been supported in part by Department of
Energy contract No.~DE-FG02-91ER40685 and NSF grant PHY-9970703.
%
%
\vskip -1.mm

%
%
%
\begin{table}
\caption{The measured CDF (preliminary) and predicted average $A_{FB}$ in the
$300-600~GeV/c^2$ range and the probability
that models can result in the observed $A_{FB}$ (3 events in
the negative hemisphere and 1 event in the positive hemisphere)
for the Standard Model and models with an extra $Z'$ boson.}
\label{Table2}
\begin{center}
\begin{tabular}{cccc}
\multicolumn{1}{c}{$M_{Z'}$($GeV/c^2$)} & \multicolumn{1}{c}{$\phi$} &
\multicolumn{1}{c}{$A_{FB}$} & \multicolumn{1}{c}{$Probability$}
\\ \hline
        $data$  &     $-$ & $-0.45\pm0.47$  & $-$    \\
        $SM$  &     $-$ & 0.612  & $2.4\%$    \\
         350  &  $60^0$ &  0.284 & $13.4\%$    \\
        350   & $173^0$ &  0.419  & $7.7\%$     \\
        500   &  $70^0$ &  0.458  & $6.3\%$     \\
        500   & $160^0$ &  0.553  & $3.7\%$     \\
\end{tabular}
\end{center}
\end{table}
\vskip -3.mm
%
%
%
\begin{table}
\caption{CDF (Preliminary) data versus Standard Model Drell-Yan cross section
$\sigma_{ee}(fb)$
for two mass bins: $300-400$ and $400-600~GeV/c^2$.
    Also shown are total
Drell-Yan cross sections when an additional $E_6$ $Z'$ boson with a mass
$M_{Z'}$ with a total width of $\Gamma_{Z'}=0.1\, M_{Z'}$ is included.
The $Z'$ di-lepton branching ratio $BR(ee)$ and
total production cross section $\sigma_{tot}~(pb)$
(with final state decays to any particles)
are calculated for the values
of $\phi$ obtained from Rosner's fits to low energy data.}
\label{Table1}
\begin{center}
\begin{tabular}{cccccc}
\multicolumn{1}{c}{$Bin$ ($GeV/c^2$)} &
\multicolumn{1}{c}{$M_{Z'}$($GeV/c^2$)} & \multicolumn{1}{c}{$\phi$} &
\multicolumn{1}{c}{$\sigma_{ee}(fb)$} & \multicolumn{1}{c}{$BR(ee)$} &
      \multicolumn{1}{c}{$\sigma_{tot}(pb)$} \\ \hline
      $300-400$& $data$  & $-$ &  $81\pm47$  & $-$        & $ -    $  \\
      $300-400$ & $SM$  & $-$ &  54  & $-$        & $ -    $  \\
      $300-400$ & 350   &  $60^0$ &  127 & $0.43\%$   & $ 17.1 $  \\
      $300-400$ & 350   & $173^0$ &  85  & $0.29\%$   & $ 10.9 $  \\
      \hline
      $400-600$& $data$  & $-$ &  $28\pm28$  & $-$        & $ -    $  \\
      $400-600$ & $SM$  & $-$ &  17  & $-$        & $  - $    \\
      $400-600$ & 500   &  $70^0$ &  34  & $0.53\%$   & $ 3.2   $  \\
      $400-600$ & 500   & $160^0$ &  24  & $0.41\%$   & $ 1.7   $  \\
\end{tabular}
\end{center}
\end{table}

\end{document}